\documentclass[10pt]{iopart}
\usepackage{graphicx}
\usepackage{xspace}				% Add spaces after aliases
\usepackage[pdftex]{color}			% Enable colored text

\newcommand{\dmu}{\textsl{DMU}\xspace}

\newcommand{\esa}{\textsl{ESA}\xspace}
\newcommand{\fee}{\textsl{FEE}\xspace}
\newcommand{\ieec}{\textsl{IEEC}\xspace}

\newcommand{\lisa}{\textsl{LISA}\xspace}
\newcommand{\lpf}{\textsl{\mbox{LPF}}\xspace}
\newcommand{\ltp}{\textsl{LTP}\xspace}
\newcommand{\mbw}{\textsl{MBW}\xspace}
\newcommand{\nasa}{\textsl{NASA}\xspace}
\newcommand{\net}{\textsl{NET}\xspace}
\newcommand{\ntc}{\textsl{NTC}\xspace}

\newcommand{\tc}{\textsl{TC}\xspace}
\newcommand{\tms}{\textsl{TMS}\xspace}
\newcommand{\upc}{\textsl{UPC}\xspace}

\begin{document}

% Journal identifier can be put here: Classical and Quantum Gravity, 6
% Journal identifier can be put here: Measurement Science an Technology: 24

\jl{6}

\title[Extension of the \ltp temperature diagnostics\ldots]{Extension 
of the \ltp temperature diagnostics to the \lisa band: first results}

\author{J~Sanju\'an$^{1,2}$, J~Ramos-Castro$^3$ and A~Lobo$^{1,2}$}

\address{$^1$ Institut de Ci\`encies de l'Espai, {\sl CSIC}, Facultat de
 Ci\`encies, Torre C5 parell, 08193 Bellaterra, Spain}

\address{$^2$ Institut d'Estudis Espacials de Catalunya (\ieec), Edifici Nexus,
Gran Capit\`a 2-4, 08034 Barcelona, Spain}

\address{$^3$Departament d'Enginyeria Electr\`onica, \upc, Campus Nord,
 Edifici C4, Jordi Girona 1-3, 08034 Barcelona, Spain}

\ead{sanjuan@ieec.fcr.es}

%\date{\today}
\date{05-October-2008}

\begin{abstract}
High-resolution temperature measurements are 
required in the \ltp, i.e., 10 $\mu{\rm K}\, {\rm Hz}^{-1/2}$ from 
1 mHz to 30 mHz. This has been already 
accomplished with thermistors and a suitable 
low noise electronics. However, the frequency 
range of interest for \lisa goes down to 0.1 mHz. 
Investigations on the performance of temperature 
sensors and the associated electronics at frequencies 
around 0.1 mHz have been performed. Theoretical 
limits of the temperature measurement system and 
the practical on-ground limitations to test them 
are shown demonstrating that $1/f$ noise is not 
observed in thermistors even at frequencies around 
0.1 mHz and amplitude levels of 10~$\mu$K\,Hz$^{-1/2}$.
\end{abstract}
\submitto{\CQG}
%
%\maketitle

\section{Introduction}
\label{introduction}
The figure of merit of \lisa related with its capability 
of detecting gravitational waves is the differential 
acceleration noise between the test masses of the satellites. This 
acceleration noise has been set to~\cite{lisasrd}
\begin{equation}
 \hspace*{-6em}
 S_{\delta a, {\rm LISA}}^{1/2}(\omega)\leq 3\times 10^{-15}\,\left\{\left[
 1 + \left(\frac{\omega/2\pi}{8\ {\rm mHz}}\right)^{\!\!4}\right]\!
 \left[1 + \left(\frac{0.1\ {\rm mHz}}{\omega/2\pi}\right)\right]
 \right\}^{\!\frac{1}{2}} \ {\rm m}\,{\rm s}^{-2}\,{\rm Hz}^{-1/2}
 \label{lisa.req.acc}
\end{equation}
in the frequency band between 0.1~mHz and 0.1~Hz.

\lisa Pathfinder (\lpf) is the pre-\lisa mission in charge of 
demonstrating the capability of putting two test masses in 
free-fall at the levels needed for \lisa. However, the \lpf 
acceleration noise budget is reduced by one order of magnitude, both 
in amplitude and in frequency band~\cite{lpfsrd}, i.e.
\begin{equation}
 %\hspace*{-6.5em}
 S_{\delta a, {\rm LPF}}^{1/2}(\omega)\leq 3\!\times\!10^{-14}\,\left[
 1 + \left(\frac{\omega/2\pi}{3\ {\rm mHz}}\right)^{\!\!2}\right]\,
 {\rm m}\,{\rm s}^{-2}\,{\rm Hz}^{-1/2}
%, \ 1\,{\rm mHz}\leq\omega/2\pi\leq30\,{\rm mHz}
 \label{lobo-eq2}
\end{equation}
for 1\,mHz$\leq\omega/2\pi\leq30$\,mHz.
% in the frequency band between 1 mHz and 30 mHz. 

This acceleration noise is the
result of various disturbances which limit the performance of the
experiment~\cite{ere2006}. The reades is referred to Lobo's contribution 
to this volume for an overview of these disturbances. One of the noise 
sources is temperature fluctuations which translate 
into acceleration noise in the test masses by different mechanisms~\cite{bola.paper}.
Those affecting directly the test masses are: (i) radiation pressure, 
(ii) radiometer effect and (iii) asymmetric outgassing; and the mechanisms 
disturbing the acceleration measurement, i.e., the 
interferometric subsystem are: (iv) temperature dependence of the index of 
refraction of optical components and (v) optical path length variations due to 
dilatation provoked by temperature changes~\cite{ow.paper}. This demands that temperature 
be monitored at different spots of the \ltp. The main reasons are:
\begin{itemize}
 \item to obtain information on the thermal behaviour of the \ltp,
 \item to identify the fraction of noise in the test masses due to thermal effects,
 \item to validate whether or not the theoretical models related to thermal effects 
are accurate.
\end{itemize}

The role of the temperature diagnostics in \lisa is still to be 
consolidated but, in principle, it will work at least as a noise cleaning tool, 
which will provide house-keeping data and will help to gravitational 
wave signal extraction. 

% The budgeted noise in the test masses due to temperature 
% fluctuations must be kept around one order of magnitude lower than 
% the foreseen total acceleration noise in the \ltp and in \lisa. Consequently, 
% temperature fluctuations must be limited in order not to 
% disturb the test masses and the acceleration measurement. 
The requirements for the temperature fluctuations in the \ltp and \lisa are\footnote{The 
baseline is not to exceed the 10\% of the total acceleration noise budget.}~\cite{bola.paper}
\begin{eqnarray}
 S^{1/2}_{T, \ LTP}(\omega)&\leq&100 \,\mu{\rm K}\,{\rm Hz}^{-1/2}, \hspace{0.3cm} 1\, {\rm mHz} \leq \omega/2\pi\leq30\, {\rm mHz} \\
 S^{1/2}_{T, \ LISA}(\omega)&\leq&10\,\mu{\rm K}\,{\rm Hz}^{-1/2}, \hspace{0.3cm} 0.1\, {\rm mHz} \leq \omega/2\pi\leq100\, {\rm mHz}
\end{eqnarray}

The temperature measurement system (\tms) must be capable of measuring the expected 
temperature fluctuations within certain accuracy, for this reason the 
Noise Equivalent Temperature (\net) must be 
about one order of magnitude lower than the expected fluctuations~\cite{fee.paper}, i.e.:
\begin{eqnarray}
 S^{1/2}_{T, \ LTP \ {\rm meas}}(\omega)&\leq&10 \,\mu{\rm K}\,{\rm Hz}^{-1/2}, \hspace{0.3cm} 1\, {\rm mHz} \leq \omega/2\pi\leq30\, {\rm mHz} \label{net.ltp} \\
 S^{1/2}_{T, \ LISA \ {\rm meas}}(\omega)&\leq&1\,\mu{\rm K}\,{\rm Hz}^{-1/2}, \hspace{0.3cm} 0.1\, {\rm mHz} \leq \omega/2\pi\leq100\, {\rm mHz}
\end{eqnarray}

The \tms of the \ltp has been already successfully tested and integrated in the 
Data Management Unit (\dmu) of the \ltp. Once we know the \tms is compliant with 
the requirements of the \ltp~\cite{fee.paper,fm.tr}, the next  natural step is to investigate the 
performance of the system at lower frequencies, i.e., extend the \tms of the \ltp to the 
\lisa band.

This article describes the noise investigations in the \tms of the \ltp. The goal 
of this investigation is to detect whether or not the \tms designed 
for the \ltp exhibits $1/f$ noise at frequencies lower than 1 mHz, specifically at 
frequencies around 0.1~mHz (\lisa lowest frequency of interest). It is organised as follows: 
in section \ref{ltp.requirement.assessment} a brief review of the results obtained with the \tms 
regarding the \ltp requirements is shown. Section \ref{extension.lisa} details the problems related 
with the tests to assess the \tms performance in the \lisa measurement bandwidth (\mbw). 
Section \ref{fb.ff.section} describes the active/passive temperature control designed to 
 carry out successful tests at 0.1 mHz, and section \ref{conclusions} shows the 
results and conclusions from the experiments.
 
\section{\tms requirement assessment for the \ltp}
%\label{tms.description}
% 
% \subsection{Design description}
% \label{tms.design.description}
% 
% The \tms for the \ltp is basically constituted by the following parts:
% \begin{itemize}
%  \item the temperature sensor: Negative Temperature Coefficient 
% 	(NTC) thermistor\footnote{BetaTherm thermistors of nominal resistance 10 k$\Omega$.},
%  \item a Wheatstone bridge with an \emph{ac} supply voltage\footnote{5~Hz square wave signal.} is used 
% 	to detect changes in the resistance of the thermistor,
%  \item the amplification stage, which consists of a instrumentation 
% 	amplifier and an anti-aliasing filter,
%  \item the analog-to-digital converter stage, a 16-bit A/D converter
%  \item the digital demodulation process, which is implemented digitally.
% \end{itemize}
% 
% The measurement chain of the \tms is shown in a block diagram in figure \ref{processing.scheme}.
% \begin{figure}[h]
% \centering
% 	\includegraphics[width=0.65\columnwidth]{./figures/processing.pdf}
% \caption{Block diagram of the \tms of the \ltp. The upper part corresponds to 
% the analog signal processing while the diagram in the bottom reflects the 
% digital demodulation. $x(t)$ and $m(t)$ are the voltage output and the modulating 
% square wave signal, respectively. \label{processing.scheme}}
% \end{figure}

\begin{figure}[h]
 \centering
	\includegraphics[width=0.3\columnwidth]{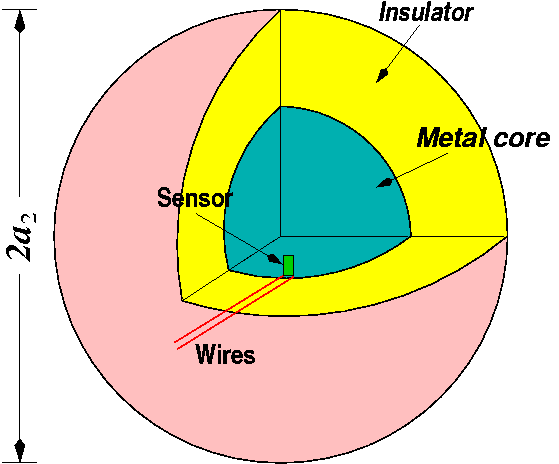}
\hspace{0.5cm}
 	\includegraphics[width=0.45\columnwidth]{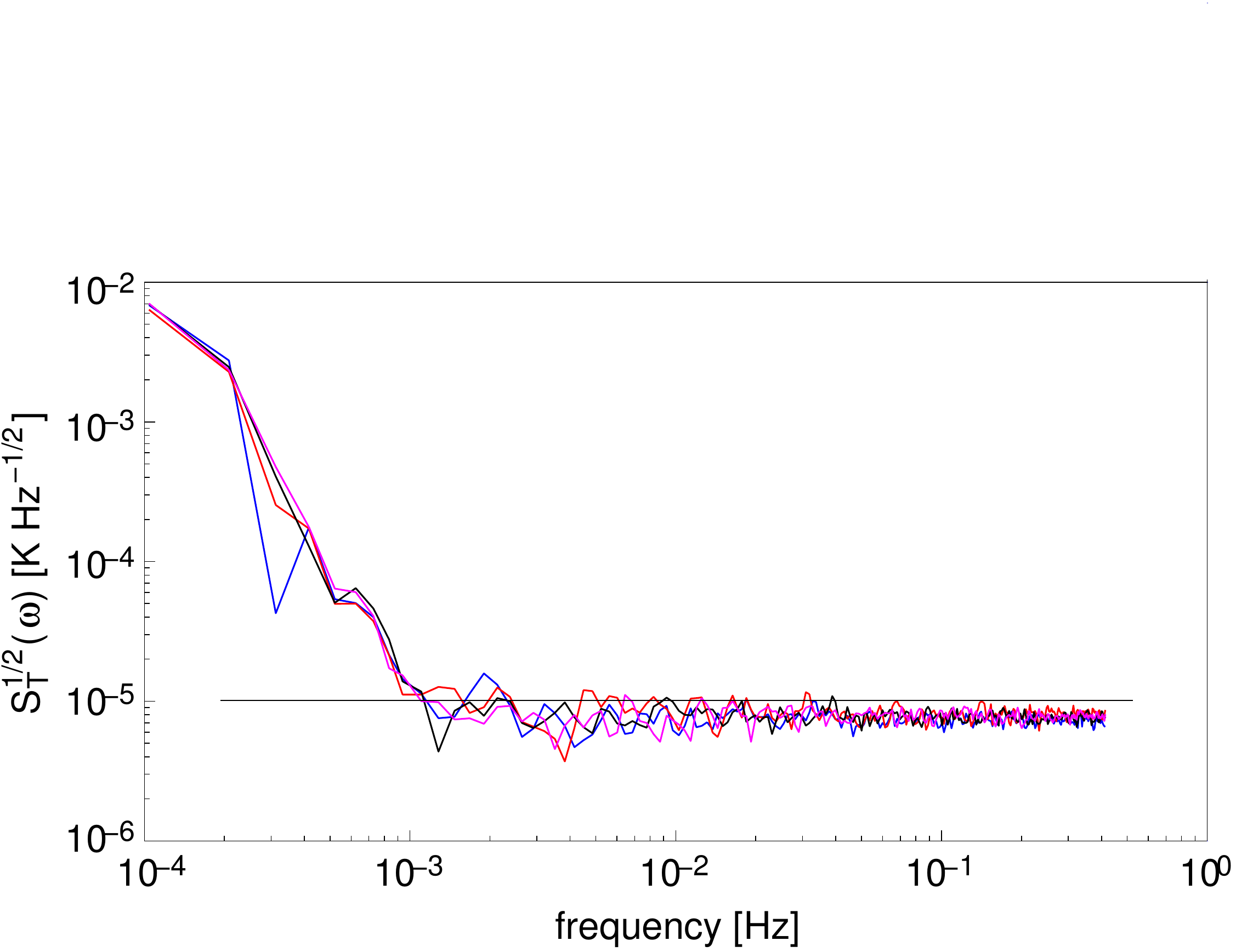}
 \caption{Left: Thermal insulator concept. A block of aluminium (0.2~m diameter) is surrounded 
by a polyurethane foam layer (0.6~m diameter). The attenuation of this jig at 
1~mHz is enough to shield the ambient temperature fluctuations at the micro-Kelvin level.
Right: Results from the \ltp \tms flight model tests validation. \label{insulator}}
 \end{figure}

%\subsection{\tms requirement assessment for the \ltp}
\label{ltp.requirement.assessment}
The \tms implemented in the \dmu of the \ltp consists of a Negative 
Temperature Coefficient (\ntc) thermistor (BetaTherm of nominal resistance 10 k$\Omega$)
followed by a low-noise signal processing, fully detailed in~\cite{fee.paper}. 
The requirement of such system for the \ltp is 10~$\mu{\rm K}\,{\rm Hz}^{-1/2}$ for 
$\omega/2\pi\geq1$\,mHz. In order to assess whether or not 
the system is compliant with the requirement we need to place 
the thermistors in a thermal environment more stable than 
the required \net, i.e., we must screen out laboratory temperature fluctuations 
to measure only the noise of the system. This implies the need 
to construct an insulator able to shield ambient 
temperature fluctuations about 5 orders of magnitude in the mili-hertz range.
 The concept of the insulator and the results obtained for a 
test campaign are shown in figure~\ref{insulator} where it is shown that 
the requirements for the \ltp are fulfilled~\cite{bola.paper,fee.paper}.

\section{Extension to the \lisa bandwidth: 0.1\,mHz}
\label{extension.lisa}
Once the requirements for the \ltp have been accomplished the next 
question, with an eye on \lisa, is: do we have the same noise levels at 0.1~mHz? 
In principle, the electronic noise of the \tms should remain flat for all the frequencies 
due to the measurement method used~\cite{fee.paper}, however, unexpected 
excess noise might appear at very low frequencies (0.1~mHz), specially, due to 
the semiconductor nature of the temperature sensor itself~\cite{pallas}.
% The potential noise 
% of thermistors is not avoided by the \emph{ac} measurement method.

The tests in the submili-hertz region are considerably more complicated
than the tests performed for the mili-hertz range. At frequencies down to 
0.1\,mHz the ambient temperature fluctuations become
 much larger and, in turn, the passive insulator attenuation appears to be quite
poor, unless a prohibitively large insulator is built. The \emph{nominal} ambient temperature 
fluctuations in the laboratory and the required temperature fluctuations inside the insulator are
\begin{eqnarray}
 S^{1/2}_{T, \ {\rm amb}}(0.1\,{\rm mHz})&\approx& 1\ {\rm to}\ 10 \ {\rm K}\,{\rm Hz}^{-1/2} \label{lisa.ambient}\\
 S^{1/2}_{T, \ {\rm ins}}(0.1\,{\rm mHz})&\approx& 10 \ \mu{\rm K}\,{\rm Hz}^{-1/2} \label{lisa.insulator}
\end{eqnarray}
From equations (\ref{lisa.ambient}) and (\ref{lisa.insulator}) the needed attenuation at 0.1~mHz
is readily calculated
\begin{equation}
 |H_{\rm ins}(0.1 {\rm mHz})|\leq 10^{-5}\ {\rm to} \ 10^{-6}
\end{equation}

In order to obtain such an attenuation the needed passive insulator would consist 
in a aluminium sphere of 0.4~m diameter surrounded by a 2~m diameter polyurethane foam layer. 
However, the most discouraging 
reason to discard using a passive insulator is its time constant which is about 
10~days. In consequence, temperature stabilisation down to 
10~$\mu{\rm K}\,{\rm Hz}^{-1/2}$ at 0.1~mHz by using a mere passive insulator appears
unreasonable.

\subsection{Solution: differential measurements}
\label{lisa.differential}
A straight forward solution to avoid the need of a giant insulator is the 
use of differential temperature measurements instead of absolute temperature 
measurements\footnote{``Differential measurements'' stands for measuring the temperature difference 
between two thermistors 
very close to one another in a more or less thermally stable environment while ``absolute measurements'' stands for measuring the temperature of individual thermistors.}. For 
the differential measurements we have that
\begin{eqnarray}
 S_{T_{1}}(\omega)=|H_{\rm ins}(\omega)|^{2}S_{T, {\rm amb}}(\omega) + n_{T_{1}}(\omega) \label{st1.abs} \\
 S_{T_{2}}(\omega)=|H_{\rm ins}(\omega)|^{2}S_{T, {\rm amb}}(\omega) + n_{T_{2}}(\omega) \label{st2.abs}
\end{eqnarray}
where $n_{T_{1}}$ and $n_{T_{2}}$ are the noise of each of the thermistors plus electronic noise. Thus, the differential measurement fluctuations are\footnote{Assuming the noise of the two thermistors is uncorrelated.}
\begin{equation}\label{diff.meas}
 S_{\Delta T}(\omega)=n_{T_{1}}(\omega)+n_{T_{2}}(\omega)
\end{equation}

Ideally, when differential 
measurements are made, only noise from the thermistors and the electronics is  
measured, since the common temperature fluctuations cancel out. Therefore,
a very simple passive insulator would be enough. 
However, different non-idealities 
arise in the practical implementation which
are discussed in the next section. Finally, we must keep in mind that the main concern 
about the potential $1/f$ noise in the submili-hertz range comes from the \ntc thermistors which 
can be detected despite the differential measurements 
since we assume the $1/f$ noise of both thermistors 
is uncorrelated. 

\subsection{Real world limitations}
\label{real.world.lim}

Different non-idealities impede the straightforward tests of the 
\tms at very low frequency proposed in the previous section. 
The main limitations in the 
measurement that need to be overcome are:
\begin{itemize}
 \item temperature coefficient (\tc) of the electronics ($K_{FEE}$),
 \item cables connecting the thermistors to the electronics ($K_{\rm C}$),
 \item intrinsic differences between thermistors ($K_{NTC}$).
 %\item thermal resistance related 
\end{itemize}

All these effects couple into the measurement and 
disturb the $1/f$ noise investigations. The apportioning of the 
these effects appears in the differential temperature measurements 
as (we have omitted the frequency dependent argument in the notation for simplicity)
\begin{equation}
 S_{\Delta T}\approx n_{T_{1}}+n_{T_{2}}+K^{2}_{FEE}S_{T, \, {FEE}} + K^{2}_{\rm C}S_{T, \, {\rm amb}} + K^{2}_{NTC}S_{T, \, {\rm ins}}
\end{equation}
where $S_{T, \, FEE}$, $S_{T, \ {\rm amb}}$ and $S_{T, \ {\rm ins}}$ stand for the 
temperature fluctuations in the electronics, in the laboratory and inside the insulator, respectively. 
The main objective is to determine $n_{T_{1}}+n_{T_{2}}$, hence, all the 
other contributions must be minimised. More specifically, we assign $\approx5\mu{\rm K}\,{\rm Hz}^{-1/2}$ for each of the three disturbing terms.

The following sections detail these non-idealities and the solution 
adopted to overcome them.

\subsubsection{Electronics temperature coefficient}
\label{temperature.coeff}
The \tc of the electronics implies 
that its temperature fluctuations appear 
in the temperature read-out as a \emph{fake} temperature~\cite{fee.paper}, i.e.,
\begin{equation}
 S_{\Delta T}(\omega)=K^{2}_{FEE}S_{T, {FEE}}(\omega)
\end{equation}
% where $K^{2}_{\rm FEE}$ is the global temperature coefficient\footnote{Assumed frequency independent.} 
% of the \fee and $S_{T, {FEE}}$ is the temperature fluctuation of the electronics.

\begin{figure}[h]
 \centering
	 \includegraphics[width=0.6\columnwidth]{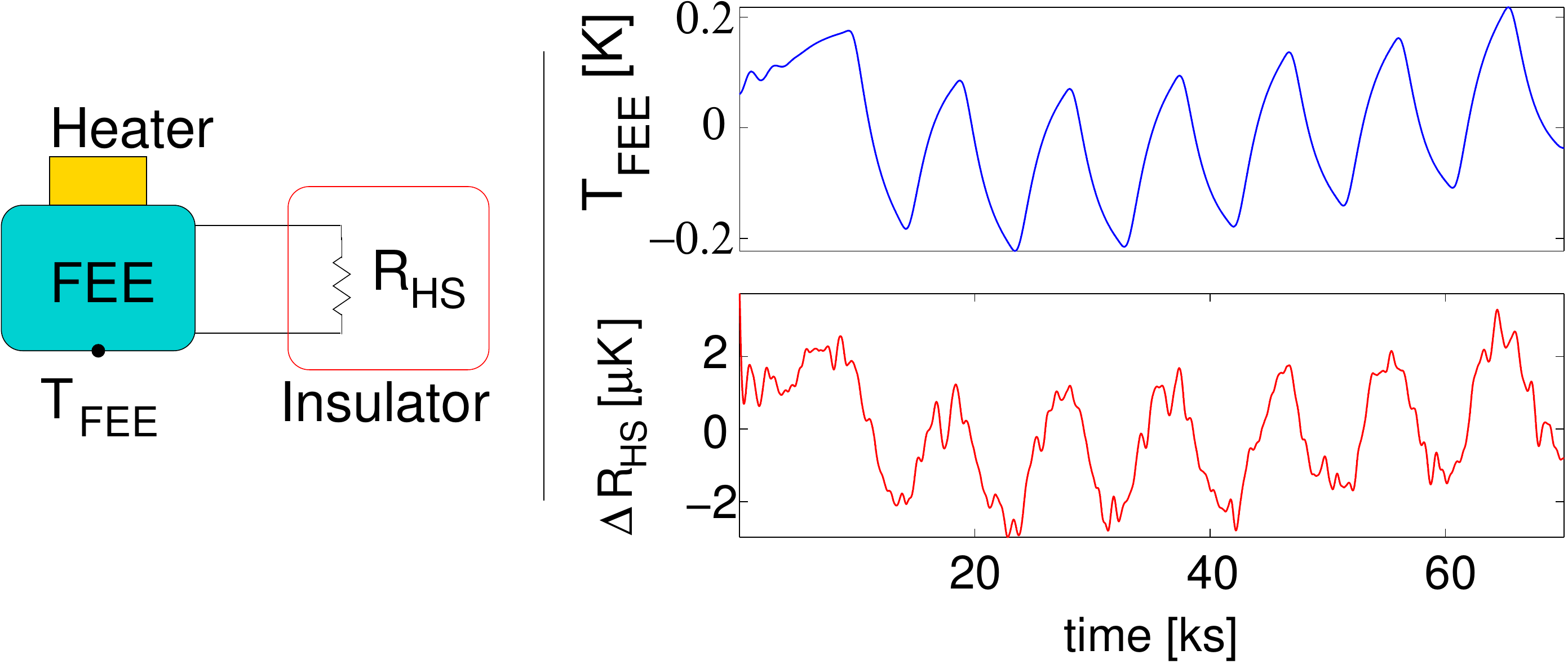}
 \caption{Left: test set-up scheme for the \tc determination. One heater is used to excite 
thermally the \fee while its temperature and a high-stability resistor are measured. Top right: measured
temperature of the \fee. Bottom right: high-stability resistance measurements in equivalent temperature. \label{TC.test}}
 \end{figure}

To determine the \tc a high-stability resistor\footnote{10 k$\Omega$ Vishay resistor with \tc of $0.6\times10^{-6}$~K$^{-1}$ placed inside the insulator.} instead of 
a thermistor was connected to the \fee. The \fee was thermally excited by using a heater 
in order to estimate the \tc when measuring the high-stability resistor.
A scheme of the test set-up and the obtained measurements are shown in figure \ref{TC.test}.

From the measurements shown in figure \ref{TC.test} (right) the estimation of the parameter $K_{FEE}$ can be easily done. The value obtained is 
\begin{equation}
 \widehat{K}_{FEE}=13.5 \ \mu{\rm K}/{\rm K}
\end{equation}
This implies that in order to keep the effect of the \tc below 5~$\mu$K\,Hz$^{-1/2}$ 
the \fee temperature fluctuations must be lower than 0.35~K\,Hz$^{-1/2}$ within the \textsl{MBW} which, 
in turn, means that the \fee must be thermostated since the ambient temperature fluctuations 
are around 1~K\,Hz$^{-1/2}$ at the 0.1~mHz range\footnote{It is considered that the \fee temperature fluctuations are the same as the laboratory ones.}. The solution to this problem was solved by 
controlling the \fee temperature by means of a feedback control ---see figure~\ref{fbff.impl} (right).

\subsubsection{Cables connecting thermistors to the \fee}
\begin{figure}[h]
 \centering
         \includegraphics[width=0.65\columnwidth]{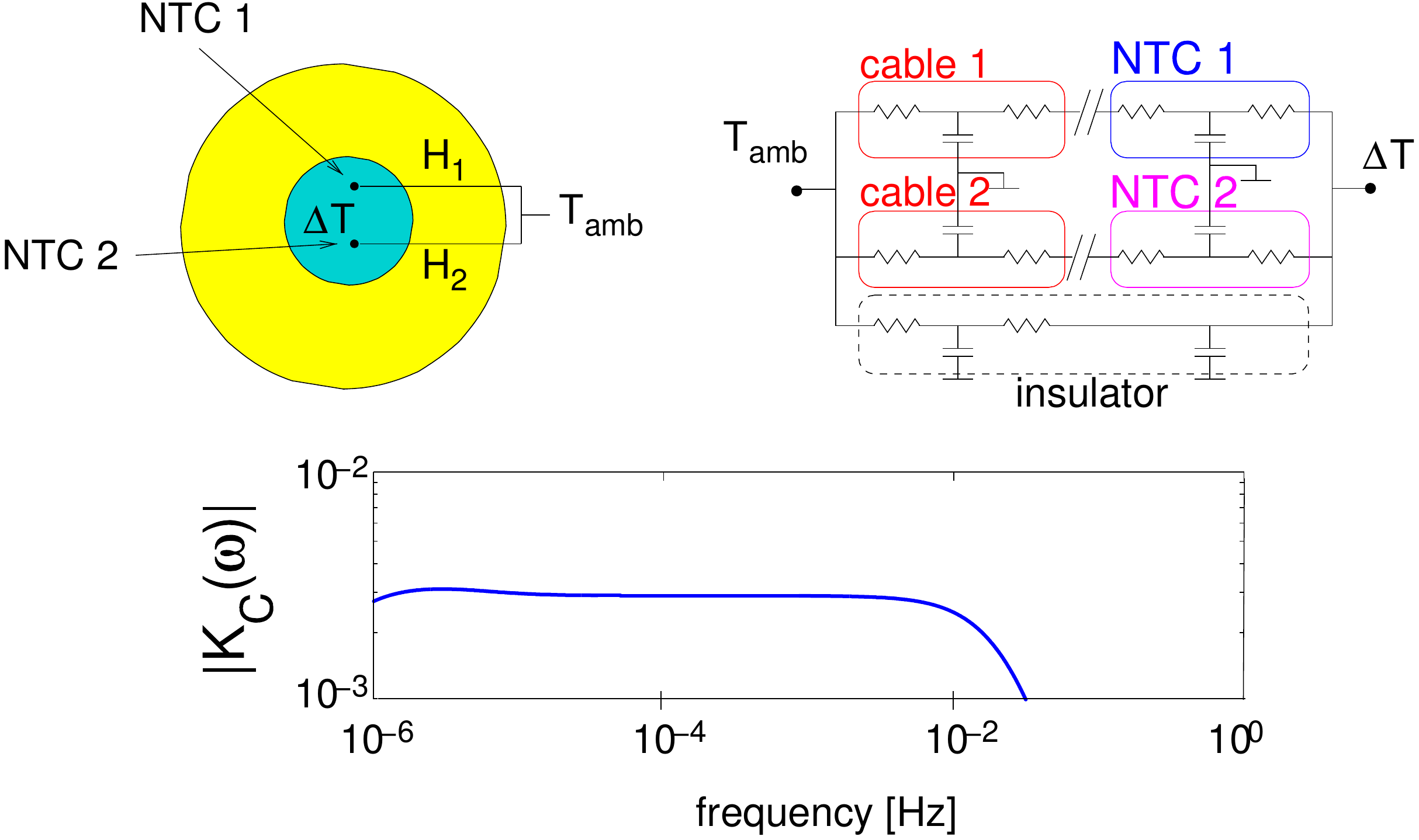}
 \caption{Common mode ambient temperature is not fully rejected due to mismatching between $H_{1}$ and $H_{2}$. These transfer functions include cables and thermal contact properties. Left: Lumped model for the $H_{1}-H_{2}$ transfer function. Right: $H_{1}-H_{2}$ transfer function. Cables are assumed identical for both sensors ($\ell$=0.25~m and 24 AWG) and thermal resistance are (measured): 95~K\,W$^{-1}$ and 100~K\,W$^{-1}$ for each of the thermistors.\label{cables.test}}
 \end{figure}

Cables are a fair way for the ambient temperature fluctuations to appear in the 
thermistors' measurements~\cite{bola.paper, fee.paper}. Theoretically, differential measurements 
should fully reject the common mode ambient temperature, however, this is 
not the case due to mismatching between the cables and thermal resistance of 
both thermistors. This can be expressed as follows
\begin{equation}
 S_{\Delta T}(\omega)=|H_{1}(\omega)-H_{2}(\omega)|^{2}S_{T,\ {\rm amb}}(\omega)
\end{equation}
where $H_{1}$ and $H_{2}$ are the transfer functions relating the ambient temperature 
fluctuations to the thermistors' ones. An estimation of $H_{1}-H_{2}$ was obtained by modeling 
the cables and the thermal contacts using a electrical analogy~\cite{incropera} ---see figure~\ref{cables.test}.

% In order to obtain a estimation of $H_{1}-H_{2}$ a electrical analogy
% has been used to model the cables and the thermal contact. The model used is shown in 
% figure \ref{cables.model} (left). The transfer function obtained is given in figure \ref{cables.model} (right).
%  \begin{figure}[h]
%  \centering
% 	 \includegraphics[width=0.45\columnwidth]{./figures/cables_model.pdf}
% 	 \includegraphics[width=0.45\columnwidth]{./figures/cables_tf.pdf}
%  \caption{Left: Lumped model for the $H_{1}-H_{2}$ transfer function. Right: $H_{1}-H_{2}$ transfer function. Cables are assumed identical for both sensors ($\ell$=0.25~m and 24 AWG) and thermal resistance are: 95~K\,W$^{-1}$ and 100~K\,W$^{-1}$ for each of the thermistors. \label{cables.model}}
%  \end{figure}

\begin{figure}[h]
 \centering
	 \includegraphics[width=0.28\columnwidth]{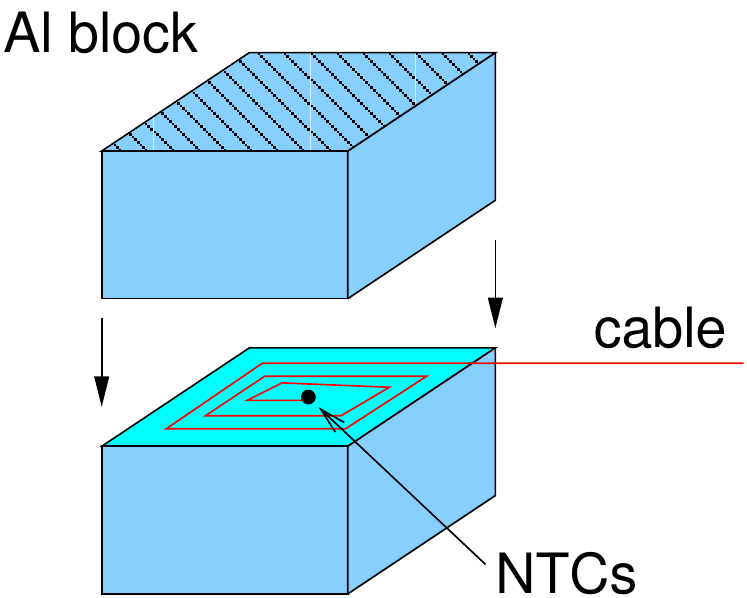} \hspace{0.5cm}
         \includegraphics[width=0.25\columnwidth]{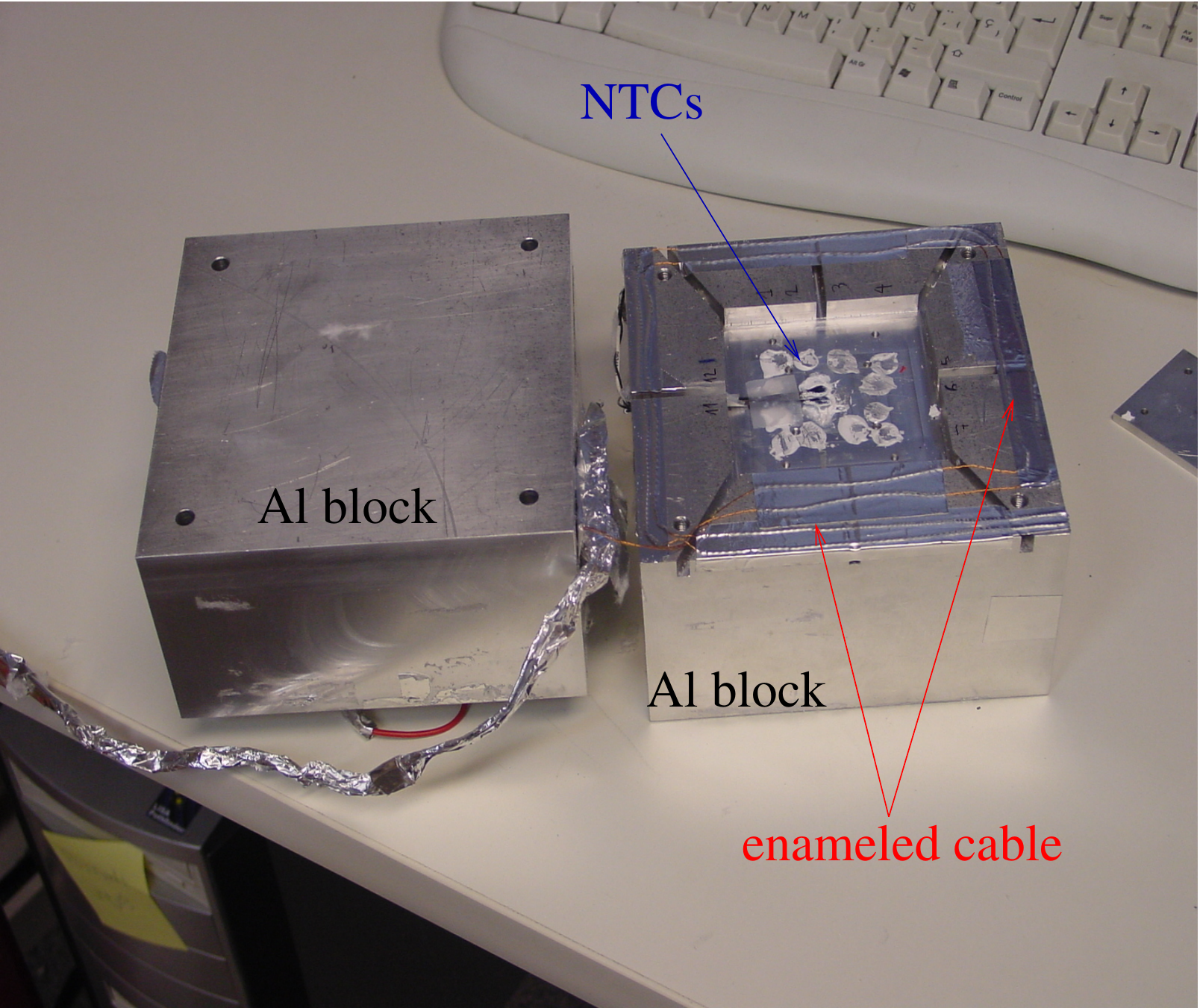}
 \caption{Left: Thermal trap concept. The ambient temperature fluctuations are attenuated 
in the aluminium block prior to \emph{reach} the sensing head of the thermistor. 
Right: Physical implementation. \label{thermal.trap}}
 \end{figure}

The rejection of the ambient temperature is $\approx3\times10^{-3}$ at the frequencies of interest. This 
means that if ambient temperature fluctuations are about 1~K\,Hz$^{-1/2}$ at 0.1~mHz, the
differential temperature measurement can result in about 3~mK\,Hz$^{-1/2}$, which is unacceptable 
considering that $\approx$5~$\mu$K\,Hz$^{-1/2}$ is the budgeted fluctuation for this effect.  
The solution adopted to reduce the common mode leakage is shown in figure~\ref{thermal.trap} and consists of~\cite{old.paper}:
\begin{itemize}
 \item use of very thin cables (AWG32 instead of AWG24),
 \item use of longer cables (from $\ell$=0.25~m to $\ell$=2~m),
 \item attach cables to the aluminium block in order to dissipate the ambient temperature fluctuations
	prior to \emph{reach} the thermistors: thermal trap.
\end{itemize}

\subsubsection{Intrinsic differences between thermistors}
Once the problem of the leakage from the ambient temperature 
is solved, there is still another non-ideality: the inherent 
mismatch between each individual 
thermistor, i.e., even when two thermistor are measuring the 
same temperature, the differential temperature read-out 
is not zero due to differences in the response 
of each thermistor. 
%Figure \ref{cables.mismatch.sch} shows this.
%  \begin{figure}[h]
%  \centering
% 	\includegraphics[width=0.25\columnwidth]{./figures/cables_mismatch_sch.pdf}
%  	\caption{The differential temperature measurement is not zero 
% even if the two sensors are placed in the same point due to the inherent mismatch between 
% both sensors. \label{cables.mismatch.sch}}
%  \end{figure}
The apportioning of this effect into the differential measurement 
is:
\begin{equation}
 S_{\Delta T}(\omega)=|H_{1}(\omega)-H_{2}(\omega)|^{2}S_{T, \ {\rm ins}}(\omega)
\end{equation}
where here $H_{1}-H_{2}$ is\footnote{Assuming the head of the thermistor is spherical. Moreover we neglect 
the different temperature coefficient, $\beta$, of the \ntc's, since this effect becomes negligible in the \mbw. }
\begin{equation}\label{mismatch_i}
 H_{1}(\omega)-H_{2}(\omega)=\frac{q_{1}}{\sinh{aq_{1}}}-\frac{q_{2}}{\sinh{aq_{2}}}
\end{equation}
with $a$ the radius of the \ntc and 
\begin{equation}\label{mismatch_ii}
 q_{j}=\frac{\rho_{j}c_{j}}{\kappa_{j}}i\omega
\end{equation}
where $\rho_{j}$, $c_{j}$ and $\kappa_{j}$ are the density, the specific heat and 
the conductivity of the thermistors and their encapsulations, respectively. Equation (\ref{mismatch_i}) is difficult to evaluate 
due to the non-accurately known characteristics of the thermistors. However, it can be seen 
that equation (\ref{mismatch_i}) is, actually, a high-pass filter, i.e.
\begin{equation}
 H_{1}(\omega)-H_{2}(\omega)\approx k_{NTC}i\omega
\end{equation}

\begin{figure}[h]
 \centering
	 \includegraphics[width=0.42\columnwidth]{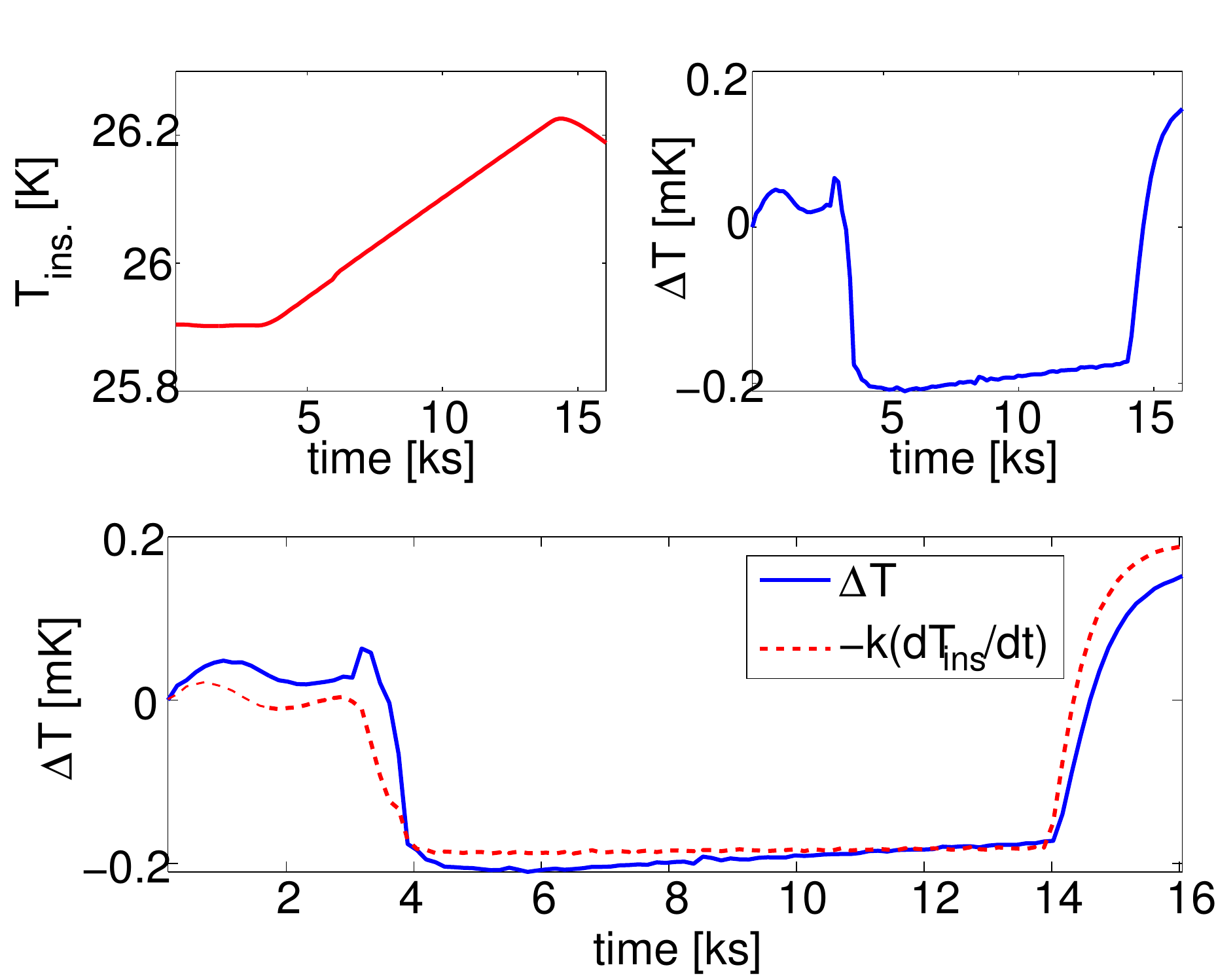}
	 \includegraphics[width=0.40\columnwidth]{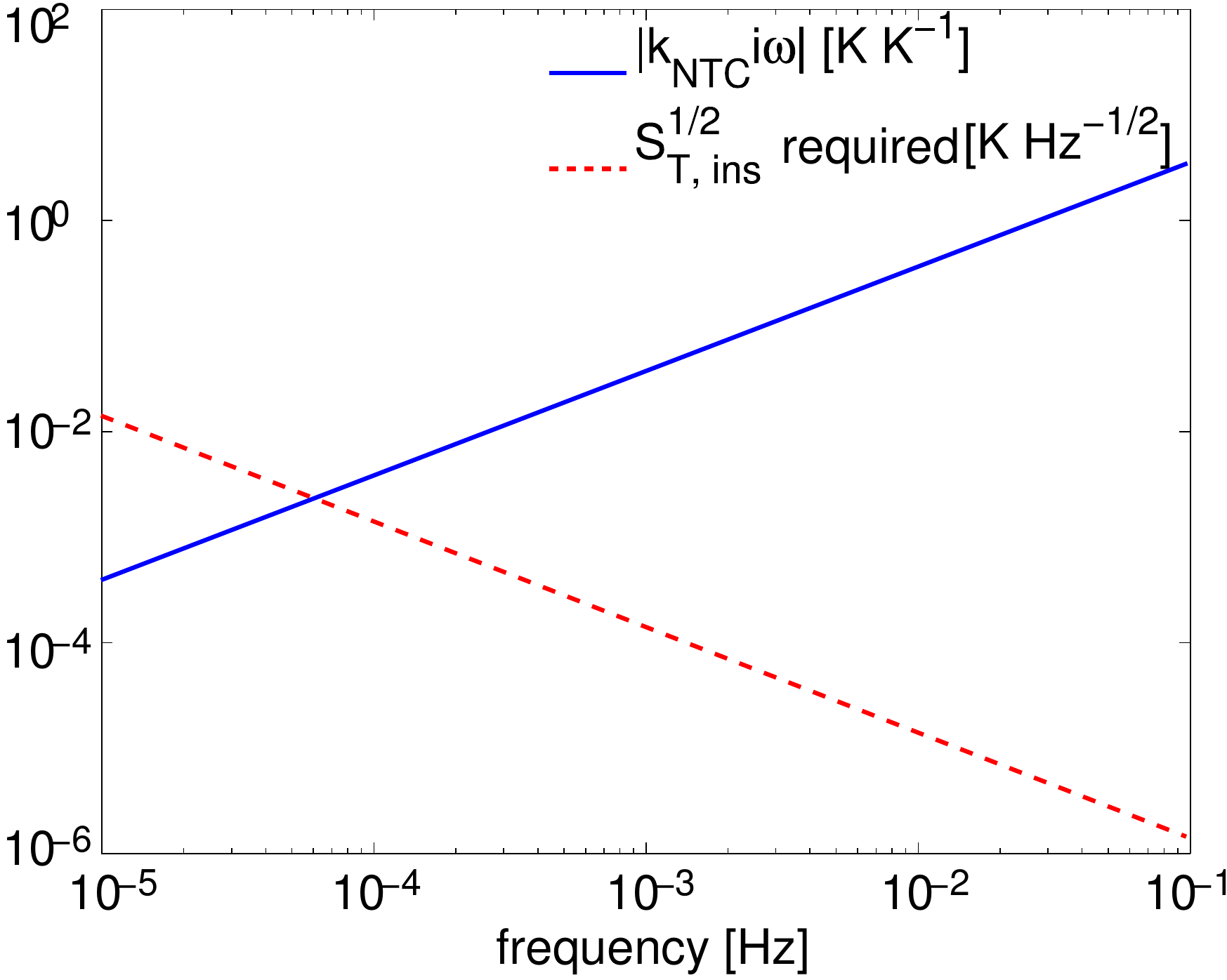}
 \caption{Left panel: the two top measurements correspond to the absolute temperature 
of the aluminium block (left) and the differential temperature measurement (right). The bottom plot 
shows the differential temperature measurement (solid trace) and the fitted one by derivation of
the absolute temperature and adjusting the parameter to $k_{NTC}$=6\,s (dashed trace). Right: solid trace is the transfer function relating absolute and differential temperature measurements. The dashed trace stands for the required temperature stability of the aluminium block to not disturb the tests. \label{cables.mismatch.test}}
 \end{figure}

Experiments to estimate $k_{NTC}$ consisted of exciting thermally 
the aluminium block and measuring the absolute and the differential temperature.
Figure \ref{cables.mismatch.test} (left) shows the measurements 
that lead to the estimation of $\widehat{k}_{NTC}$=6~s. This effect implies that the 
temperature fluctuations of the insulator, $S_{T,{\rm ins}}$, must 
be limited to $\approx5$\,mK\,Hz$^{-1/2}$ at 0.1\,mHz ---see figure \ref{cables.mismatch.test} (right)---
which is not feasible for a passive insulator. In consequence, an active 
temperature controller together with the passive insulator ---see section~
\ref{fb.ff.section}--- was needed.

\section{Active/passive temperature controller}
\label{fb.ff.section}
\begin{figure}[h]
 \centering
	 \includegraphics[width=0.55\columnwidth]{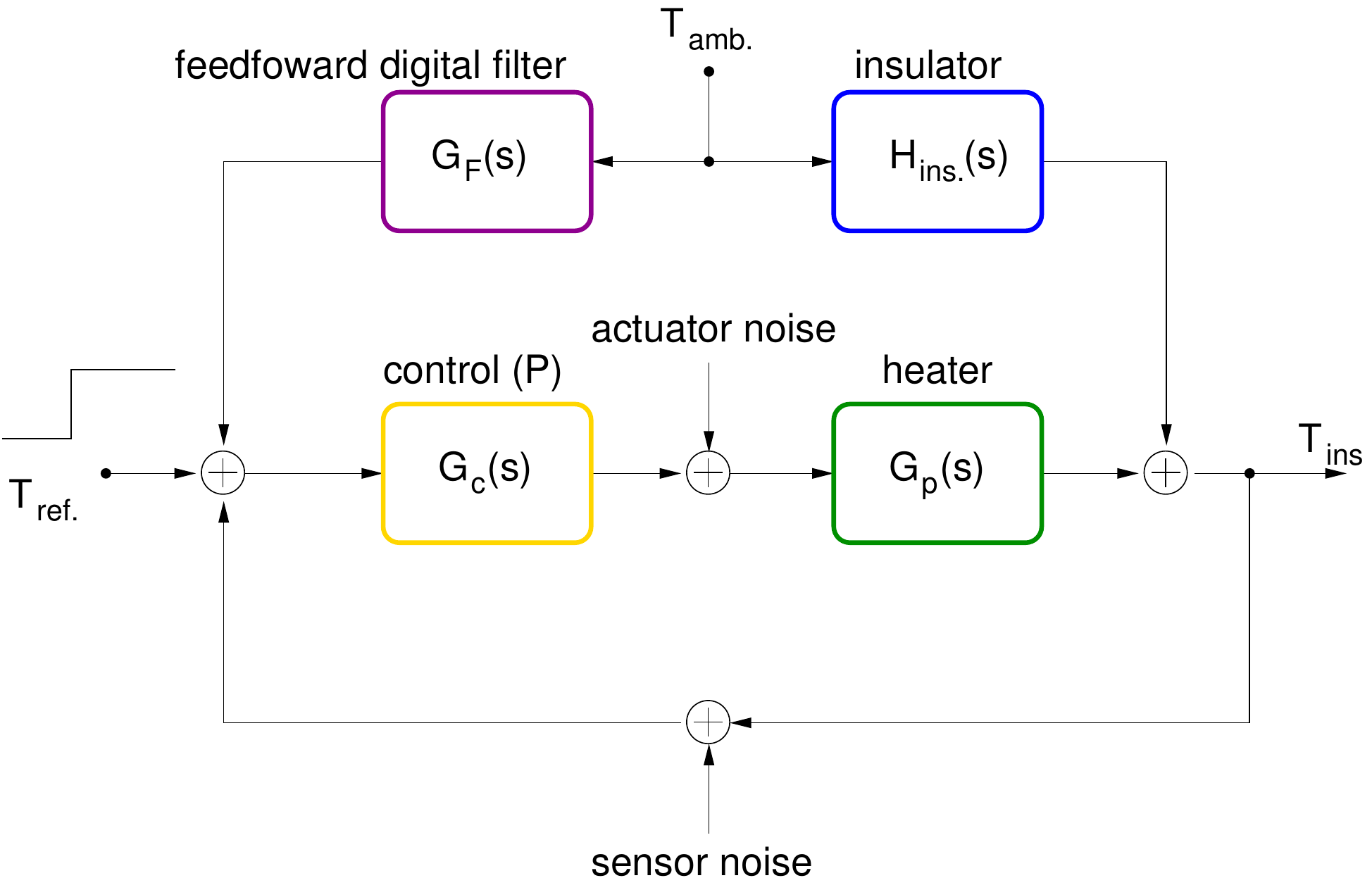}
 \caption{Feedback-feedforward control temperature system block diagram. See text for details. \label{fb.ff.sch}}
 \end{figure}

\begin{figure}[h]
 \centering
	\includegraphics[width=0.42\columnwidth]{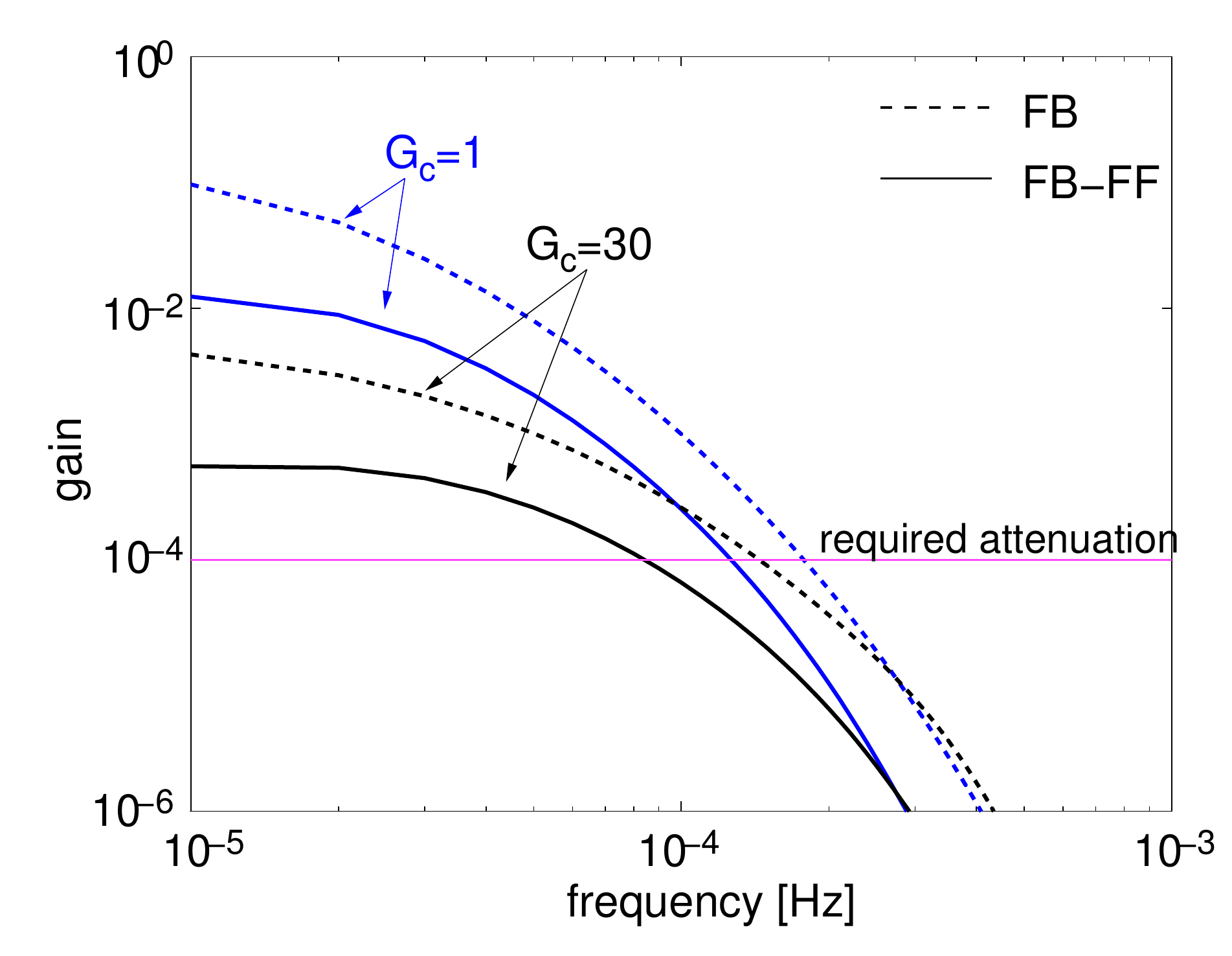}
	\includegraphics[width=0.45\columnwidth]{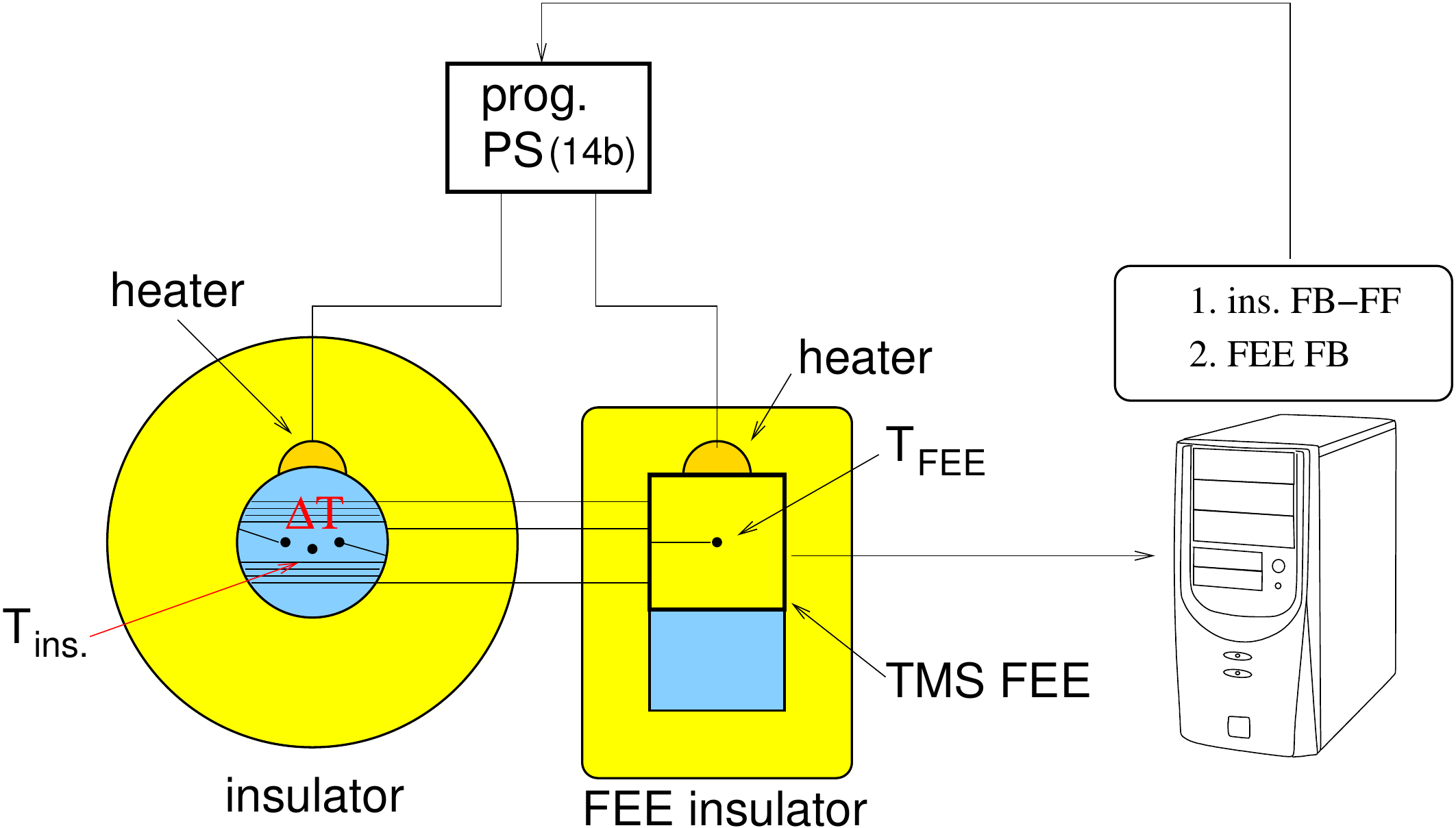}
 \caption{Left: Ambient temperature fluctuations screening. Solid traces represent the {\sl FB-FF} for different gains, and dashed traces are plotted for comparison when using a {\sl FB} scheme. Right: Control system implementation scheme. $G_{\rm C}$ was set to 30. \label{fbff.impl}}
 \end{figure}

In order to keep the temperature fluctuations of the insulator at the required
levels, a passive insulator plus an active temperature control based on a feedback-feedforward ({\sl FB-FF}) scheme have been used. The former attenuates the temperature fluctuations above the mili-hertz region 
while the latter is useful to shield ambient temperature fluctuations at the submili-hertz region. 
Since the ambient temperature can be measured, the {\sl FB-FF} control system 
can be implemented if good knowledge of the transfer functions 
involved in the control is provided~\cite{fbff.book}. The control system works as follows: a reference 
temperature for the aluminium block, $T_{\rm ref}$ (a few degrees over 
the ambient temperature), is set; then the control tries to maintain this 
temperature by dissipating power through a heater attached to the aluminium 
block. The applied power dissipated is calculated by a computer using the data coming from 
the ambient temperature (feedforward) and from the 
aluminium block temperature (feedback). The needed power is converted into a voltage which is supplied 
by a programmable power supply to the heater.
The block diagram of the control system is given in figure \ref{fb.ff.sch} and its 
implementation scheme is shown in figure~\ref{fbff.impl} (right).

The closed-loop response is (in the $s$-domain, we omit the $s$ dependence argument in the rest of the paper for notation simplicity)
\begin{eqnarray}
 T_{\rm ins}&=&\frac{G_{\rm C}G_{\rm P}}{1+G_{\rm C}G_{\rm P}}T_{\rm ref}+\frac{H_{\rm ins}-G_{\rm C}G_{\rm P}G_{\rm F}}{1+G_{\rm C}G_{\rm P}}T_{\rm amb}+ \nonumber \\
&&+\frac{G_{\rm P}}{1+G_{\rm C}G_{\rm P}}n_{\rm act}+\frac{G_{\rm C}G_{\rm P}}{1+G_{\rm C}G_{\rm P}}n_{\rm sens} \label{closed.loop}
\end{eqnarray}
where:
\begin{itemize}
 \item $G_{\rm P}$ is the heater transfer function, i.e., the transfer function that relates the 
power dissipated in the heater and the increase of the temperature of the aluminium block,
 \item $H_{\rm ins}$ is the passive insulator transfer function, i.e., the attenuation 
of the ambient temperature fluctuations provided by the passive insulator,
 \item $G_{\rm C}$ is the controller transfer function, in our case a mere constant,
 \item $G_{\rm F}$ is the feedforward filter. In order to null the term multiplying the ambient temperature 
in equation (\ref{closed.loop}) this filter must be
	\begin{equation}\label{ff.filter}
	 G_{\rm F} = \frac{1}{G_{\rm C}}\frac{H_{\rm ins}}{G_{\rm P}}
	\end{equation}
 where all the transfer functions in the rhs of the equation are known\footnote{Different tests/analyses were performed in order to obtain the transfer functions $H_{\rm ins}$~\cite{bola.paper} and $G_{\rm P}$ accurately.}. This filter is implemented digitally as a 11th order filter\footnote{The filter consists of a cascade of low-pass filters and all-pass filters in order to approximate as much as possible the gain and the phase of the desired filter ---equation~(\ref{ff.filter}).},
 \item $T_{\rm ref}$, $T_{\rm ins}$ and $T_{\rm amb}$ are the set point temperature for the aluminium block, the measured temperature of the aluminium block and the laboratory temperature, respectively,
 \item $n_{\rm act}$ and $n_{\rm sens}$ are the noise introduced by the programmable power supply and the noise 
of the absolute temperature sensor. Both are low enough not to disturb the test purposes.
\end{itemize}

Our main concern is to screen the ambient temperature fluctuations down to $\approx$5\,mK\,Hz$^{-1/2}$ at 0.1\,mHz ---see figure~\ref{cables.mismatch.test} (right, dashed trace). In consequence, the figure 
of merit of the control loop is the transfer function relating the ambient temperature and the aluminium 
block temperature, i.e., 
\begin{equation}\label{ff.filter2}
 T_{\rm ins}=\frac{H_{\rm ins}-G_{\rm C}G_{\rm P}G_{\rm F}}{1+G_{\rm C}G_{\rm P}}T_{\rm amb}
\end{equation}

This transfer function is plotted in figure~\ref{fbff.impl} (left) for different values of $G_{\rm C}$ where we realise that a value of $\approx30$ is needed for $G_{\rm C}$ in order to obtain the required attenuation at the 
submili-hertz region (higher gains lead to an oscillating temperature response). Ideally, the solid traces in figure~\ref{fbff.impl} (left) 
should all be zero, however, the digital implementation of the feedforward filter is only an 
approximation of the desired one, therefore, the differences between the approximated and the desired 
filter lead to a non-zero numerator in equation~(\ref{ff.filter2}).

\section{Results and conclusions}
\label{conclusions}
\begin{figure}[h]
 \centering
	\includegraphics[width=0.6\columnwidth]{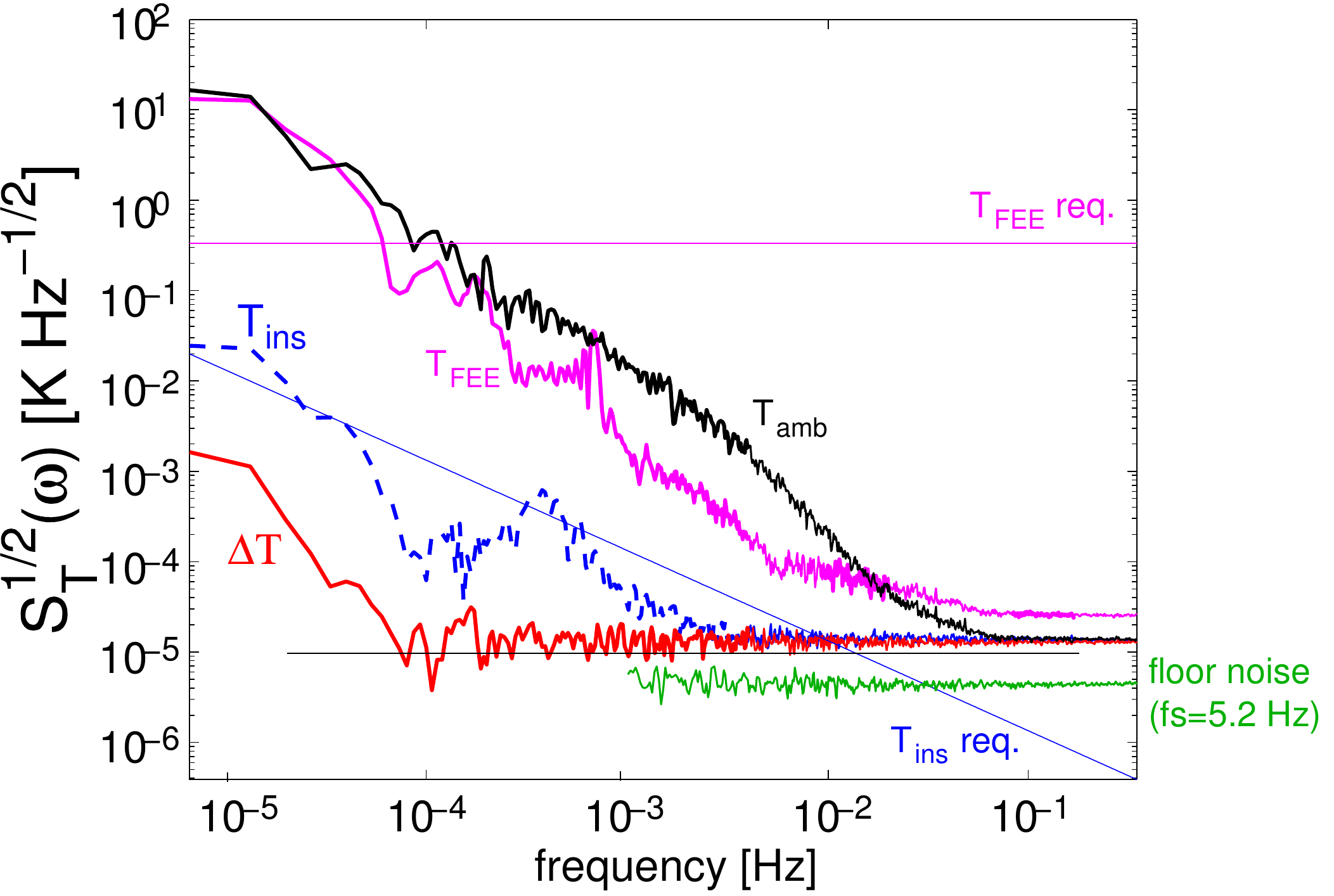}
 \caption{Tests results. ``$T_{\rm amb}$'' trace: ambient temperature fluctuations. ``$T_{FEE}$'' trace
stands for the temperature fluctuations of the \fee while the ``$T_{FEE}$ req.'' trace represents its required stability. ``$T_{\rm ins}$'' trace stands for the aluminium temperature fluctuations, the ``$T_{\rm ins}$ req.`` trace represents its
required stability. ''$\Delta T$`` trace: differential measurement temperature fluctuations where we realise that 
exceed noise is not present down to 0.1 mHz. ''floor noise`` trace: independent measurement when acquiring only one channel. The rest of measurements were done acquiring 4 channels. \label{results.plot}}
 \end{figure}

The results obtained are shown in figure~\ref{results.plot}. The solid trace labeled as ``$\Delta T$'' is the 
one standing for the differential measurement noise levels which does not exhibit any $1/f$ noise at the level of $\approx10$~$\mu$K\,Hz$^{-1/2}$ down to 0.1~mHz. We can ascribe all this noise 
to the \tms itself (thermistors and electronics) since the required temperature fluctuations of the aluminium block (``$T_{\rm ins}$'' trace) and the \fee temperature fluctuations (``$T_{FEE}$'' trace) are within the required not to affect the measurement. We have added an extra measurement (trace at the bottom of the plot) which stands for the floor noise of the \tms when channels are not multiplexed. The noise of the \tms increases when measuring more than one channel due to the aliasing by a factor $\sqrt{N}$, with $N$ the numbers of channels~\cite{fee.paper}. The tests described here were performed acquiring 4 channels plus the control system. This implies lower sampling 
frequency which translates in extra noise in all the \mbw due to aliasing.

The conclusions of the noise investigations can be summarised:
\begin{itemize}
 \item $1/f$ noise is not present in the thermistors and the associated electronics 
designed for the \ltp, i.e., the noise remains flat down to 0.1~mHz with an amplitude level of $\approx10$~$\mu$K\,Hz$^{-1/2}$,
 \item the floor noise when measuring a single channel goes down to $\approx$4\,$\mu$K~Hz$^{-1/2}$, at least, 
in the mili-hertz region,
 \item noise can be still further reduced by slight modifications in the electronics to 
reach noise levels close to 1~$\mu$K~Hz$^{-1/2}$.
\end{itemize}

The above points mean that the thermal environment of the \ltp can be measured 
with a limiting noise of $\approx$10\,$\mu$K\,Hz$^{-1/2}$ down to 0.1\,mHz. In views of \lisa, 
the \tms can be slightly modified to reduce the \net to the $\mu$K~Hz$^{-1/2}$ level.

\ack
Support for this work came from Project ESP2004-01647 of Plan Nacional del Espacio of the Spanish Ministry of Education and Science (MEC). JS acknowledges a grant from MEC.

\end{document}